\documentclass[aps,pre,reprint,superscriptaddress]{revtex4-1}
\usepackage[dvipdfmx]{graphicx}
\usepackage{dcolumn}
\usepackage{bm}
\usepackage{amssymb}
\usepackage{color}
\begin{document}

\title{\centering{Ion acceleration at two collisionless shocks in a multicomponent plasma }}
\author{Rajesh Kumar}
\affiliation{ 
Graduate School of Science, Osaka University, Japan
}%
\author{Youichi Sakawa}%
 \email{sakawa-y@ile.osaka-u.ac.jp}
\affiliation{%
Institute of Laser Engineering, Osaka University, Japan
}%
\author{Takayoshi Sano}
\affiliation{%
Institute of Laser Engineering, Osaka University, Japan
}%
\author{Leonard N. K. D$\ddot{\rm o}$hl}
\affiliation{York Plasma Institute, Department of Physics, University of York, United Kingdom}%

\author{Nigel Woolsey}
\affiliation{York Plasma Institute, Department of Physics, University of York, United Kingdom}%
\date{\today}
\author{Alessio Morace}
\affiliation{%
Institute of Laser Engineering, Osaka University, Japan
}%
\begin{abstract}
Intense laser-plasma interactions are an essential tool for the laboratory study of ion acceleration at a collisionless shock.
With two-dimensional particle-in-cell calculations of a multicomponent plasma we observe two electrostatic collisionless shocks at two distinct longitudinal positions when driven with a linearly-polarized laser at normalized laser vector potential $a_0$ that exceeds 10. Moreover, these shocks, associated with protons and carbon ions, show a power-law dependence on $a_0$ and accelerate ions to different velocities in an expanding upstream with higher flux than in a  single-component hydrogen or carbon plasma. This results from an electrostatic ion two-stream instability caused by differences in the charge-to-mass ratio of different ions. Particle acceleration in collisionless shocks in multicomponent plasma are ubiquitous in space and astrophysics, and these calculations identify the possibility for studying these complex processes in the laboratory.
\end{abstract}

\maketitle
\section{Introduction}
 Collisionless shocks under ambient magnetic field are ubiquitous in space and astrophysical plasmas, and are believed to be sources for high-energy particles or cosmic-rays \cite{Sagdeev1973,Bell1978,Blandford1978,Wu1984a,Ball2013,Hoshino2001,Sakawa2016a}.
 Multiple collisionless shocks occur in plasmas associated with planetary systems \cite{Sauer1996,Shimazu2001,Mazelle2004}, where multicomponent plasmas occur as planetary material mixes with the solar wind.
In the magnetospheres of planets, such as Mars and Venus, see \citet{Bertucci2011}, multicomponent plasmas occur and ions of differing charge-to-mass ratio likely play a role.
\citet{Jarvinen2018} discuss the role of oxygen in an induced Martian magnetosphere, where oxygen is likely introduced by the past solar wind bombardment of water on the unmagnetized surface of Mars.
Multiple-reflection of solar-wind protons at the Martian bow-shock was recorded across a shock by Mars Express and described by \citet{Yamauchi2012}.
These observations and the Voyager missions, see for example \citet{Gurnett2013}, show multiple collisionless shocks are associated with planetary and stellar systems.
\citet{Borisov2016} illustrates this for Mars and Venus where the formation of a second collisionless shock, in a region of magnetic pile-up between the bow shock and ionosphere \cite{Bertucci2005}, results from the presence of planetary oxygen ions and solar wind protons. 

Collisionless shocks occur in much more extreme astrophysical systems \cite{Warren2005,Caprioli2017,Metzger2020} such as supernova remnants where a reverse shock, an inward-propagating collisionless shock, heats stellar ejecta material containing a mixture of protons and heavy ions \cite{Warren2005}.
\citet{Warren2005} observe localized regions where strong line emission of Fe and Si ions occur in the reverse-shock heated ejecta.
\citet{Yamaguchi2014} illustrate collisionless electron heating at the front of  the reverse shock caused by a cross-shock potential created by charge deflection.
Understanding of collisionless shocks and the associated particle acceleration processes in multicomponent plasmas is of general importance in space, astrophysics, and plasma physics.

While multiple collisionless shocks are expected in such systems, it is not possible to observe them because of the limited resolution of the remote sensing.
It is possible that future spatially-resolved measurements using multi-point spacecraft clusters might observe double-shock structures. \citet{Cohen2019} and \citet{Broll2018} demonstrate {\it in situ} spatially resolved proton reflections \cite{Cohen2019} and multi-ion (solar wind protons and He$^{2+}$ contamination) reflections \cite{Broll2018} from a shock in the Earth's magnetosphere with the magnetospheric multiscale (MMS) cluster.
Laboratory experiments are a unique way of obtaining spatially resolved measurements of collisionless shocks.
They can provide tests of understanding of particle acceleration in multiple collisionless shocks.
Numerical simulations by \citet{Schaeffer2020} demonstrate the formation of two collisionless shocks as a laser-ablated plasma acts as a piston pushes on a magnetized multicomponent CH plasma.
Laboratory studies show how ion separation in unmagnetized multicomponent plasma is a common occurrence \cite{Byvank2020,Rambo1994,Bellei2014,Sio2019,Rinderknecht2018a}.
As examples, \citet{Byvank2020} use merging plasma jets at oblique angles to observe ion and shock-front separation when using jets that contain a mixture of He and Ar.
\citet{Rinderknecht2018a} observe ion velocity separation in a laser-driven collisional shock generated in a
multicomponent plasma, and  ion-species separation is predicted in inertial confinement fusion experiments as a strong shock enters the fuel containing multiple-ion species \cite{Bellei2014,Sio2019}.

Continuing advances in high-intensity laser technology \cite{Danson2019} drives the development of compact, high-flux sources of energetic ions\cite{Daido2012,Macchi2013}. 
These sources may prove useful for many applications \cite{Bulanov2014a,Li2006,Roth2001}.
Among the many ion acceleration mechanisms being pursued \cite{Snavely2000,Wilks2001,Wagner2016,Scott2018,Esirkepov2004,Henig2009a,Macchi2009a,Kim2016a,Higginson2018,Yin2006,Henig2009,stark2019}, collisionless shock acceleration (CSA) of ions \cite{Denavit1992,Silva2004,Fiuza2012,Haberberger2011,Tresca2015,Zhang2015,Zhang2017,Antici2017,Chen2017a,Pak2018a,Polz2019,Ota2019,Kumar2019a} is of particular relevance to space and astrophysical shocks.
With CSA, ions located ahead of an unmagnetized electrostatic (and collisionless) shock \cite{Sakawa2016a} are reflected by the electrostatic potential of the shock to twice the shock velocity \cite{Fiuza2012}.

Unmagnetized electrostatic collisionless shocks \cite{Forslund1970, Forslund1971} are rare among space \cite{Balogh2013} and astrophysical systems, since shocks occur in collisionless magnetized plasma. However, there are common and important collisionless processes involved in both type of shocks \cite{Silva2004}.  
For example, particle acceleration occurs in collisionless shocks \cite{Drury1983,Jones1991}, reflected particles excite two-stream instabilities \cite{Ohira2008,Treumann2009}, reflected ions cause shock dissipation and reformation \cite{Treumann2009,Balogh2013,Madanian2020}, effects of cross-shock electrostatic potential \cite{Bale2007,Cohen2019}, and so on. Bale {\it et al.} \cite{Bale2005} describe shock dissipation due to ion reflection in terms of the Cluster satellite mission.
Therefore, understanding of collisionless shocks and the associated particle acceleration processes in multicomponent plasmas is of general importance in space and astrophysical shocks.
The study of collisionless shocks and particle interaction is possible with laser-plasma systems. In this work the colllisionless shock is mediated by an electrostatic interaction.

In laser-plasma experiments hydrogen and carbon are ever-present on the surfaces of solid targets and inevitably result in multicomponent plasmas.
A number of studies \cite{Zhang2017,Antici2017,Pak2018a,Ota2019} specifically use multicomponent thin-foil targets such as plastic (CH) or Mylar ($\rm C_{10}H_8O_4$), and in \citet{Kumar2019a} we reported on how target composition influences CSA by comparing C$_2$H$_3$Cl, $\rm CH$, $\rm He^3 H$, and H.
Inclusion of a high atomic-number element like Cl, results in partial ionization, to Cl$^{15+}$, enabling the study of  a material with  $\langle Z \rangle / \langle A \rangle  <  0.5$.
An electrostatic ion two-stream instability (EITI) excited in the multicomponent plasma is central to the ion acceleration process with CSA accelerating protons \cite{Kumar2019a} and heavier ions to the same velocity.

In comparison, the radiation pressure acceleration studies by \citet{Zhang2009}, which use circularly-polarized laser pulses and a three-layer ``sandwich'' target containing protons and heavier-ions, show the emergence of two shock fronts. One shock is associated with protons and the other with  heavier ions, the different species of ion are accelerated in different fields to different velocities.
%

In this paper we examine, using the two-dimensional (2D) particle-in-cell (PIC) simulation code EPOCH \cite{Arber2015}, the physical conditions for the appearance of collisionless shocks and ion acceleration in a multicomponent plasma formed from C$_2$H$_3$Cl and $\rm CH$ targets.
We use a linearly p-polarized laser pulse, and for a normalized vector potential $a_0 \geq 10$ to show the existence of two collisionless shock fronts. These shocks are associated with the proton and C$^{6+}$ ion populations. The shock front accompanying the proton population propagates faster than the shock accompanying the C$^{6+}$ ions. As a result, CSA of protons and C$^{6+}$ ions occurs at different shocks and longitudinal locations in the plasma, producing ion populations at different velocities.

\section{Particle-in-cell simulation}
\begin{figure}
   \includegraphics[width=0.49\textwidth]{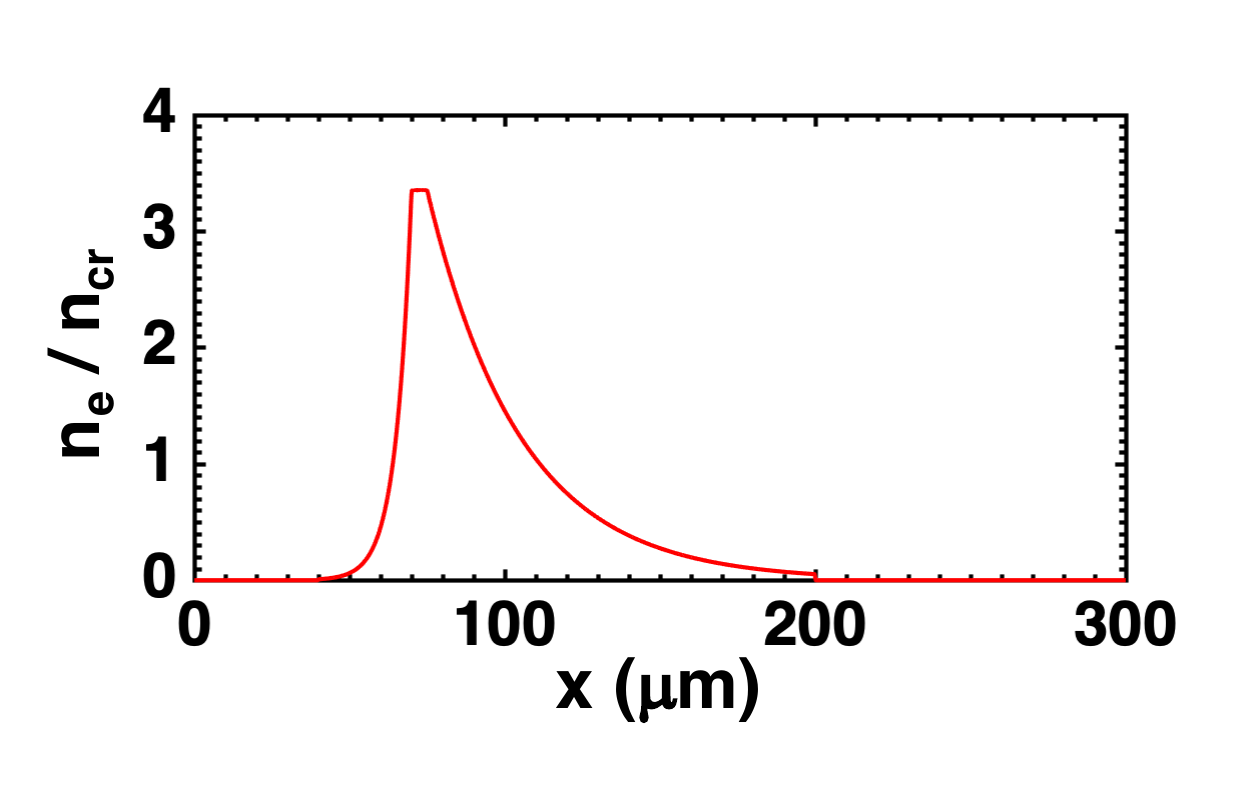}
   \caption{The normalized initial electron density profile used in PIC simulations for $a_0 = 3.35$. The laser is from the left-hand side of the simulation box. The density profile consists of an exponentially increasing $5 ~\mathrm{\mu m}$ scale-length laser-irradiated front region, followed by $5 ~\mathrm{\mu m}$ uniform central region, and an exponentially decreasing rear-side profile with $30 ~\mathrm{\mu m}$ scale-length. To avoid boundary effects, the simulations use 40 $\mu$m and 100 $\mu$m vacuum regions at the front and rear of the target, respectively.}
     \label{fig:1}
\end{figure}

\begin{figure*}
   \includegraphics[width=0.9\textwidth]{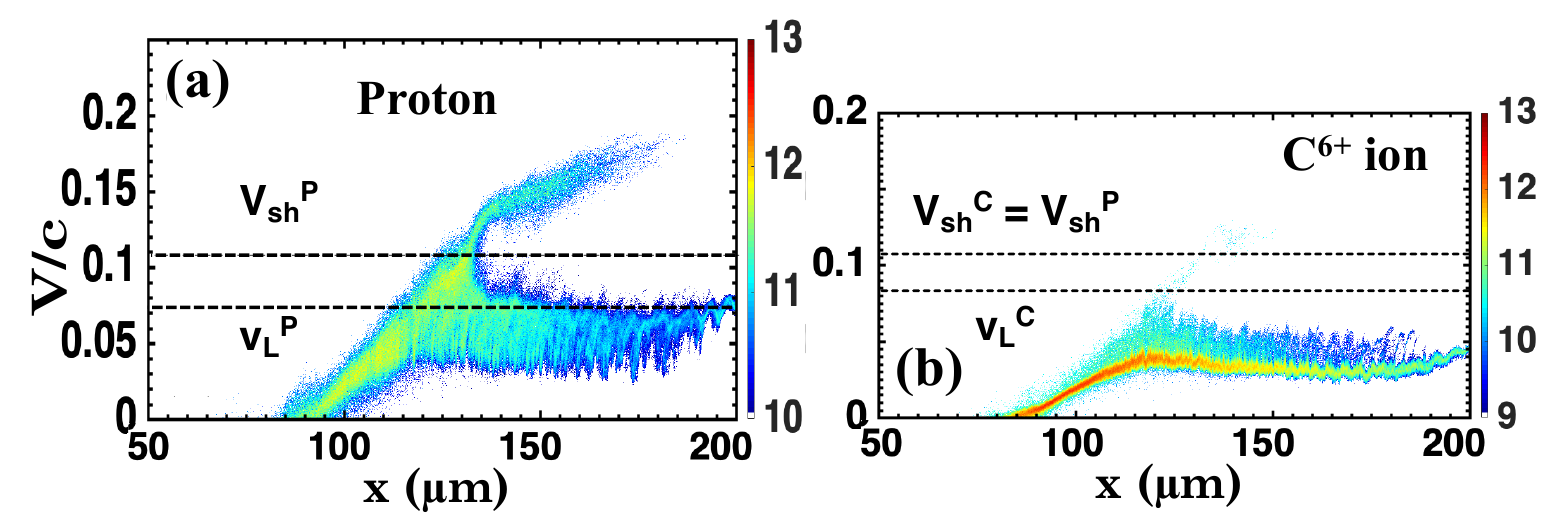}
   \caption{Phase-space of (a) protons and (b) C$^{6+}$ ions for $\rm C_2H_3Cl$ plasma at $a_0$ = 3.35 and at \textit{t} = 4.0 ps. The horizontal lines represent the lower threshold $v_L^i$ for ion reflection and the shock velocity in proton density $V_{sh}^{\rm P}$. The color scale shows the number of ions in a log scale.}
     \label{fig:11}
\end{figure*}

We study four values:  $a_0 = 3.35$, $10$, $20$, and $33$, where $a_0 = 3.35$ corresponds to $1.4 \times 10^{19}~\mathrm{W/cm^2}$ for the wavelength of $1 ~\mathrm{\mu m}$. The simulated laser pulse uses a Gaussian temporal profile with $1.5 ~\mathrm{ps}$ full-width-at-half-maximum. 
Figure \ref{fig:1} shows the normalized initial electron density profile used in PIC simulations for $a_0 = 3.35$.
The simulated targets use a longitudinal ($x$-direction in Figs.~\ref{fig:11} to \ref{fig:CH}) density profile consisting of an exponentially increasing $5 ~\mathrm{\mu m}$ scale-length laser-irradiated front region, $5 ~\mathrm{\mu m}$ uniform central region, and an exponentially decreasing profile with $30 ~\mathrm{\mu m}$ scale-length rear region as the back of the target. Details of the simulations including the target density profiles at $a_0 = 3.35$ are given in \citet{Kumar2019a}. When $a_0$ is varied, the maximum electron density is increased to match the relativistic critical density $a_0 n_{cr}$, where $n_{cr} = 1.12\times 10^{21} ~\mathrm{cm^{-3}}$ is the critical plasma density to the laser at  $1~\mathrm{\mu m}$.
The charge states $Z_i$ of protons, C-ions, and Cl-ions are  $1$, $6$, and $15$, respectively.
The corresponding ion density for each material is calculated from the quasineutral plasma condition.
%
%

At $a_0 = 3.35$, as the relativistic electrons move through the plasma, the inertia of the more massive ions sets up an electrostatic field, $E_x$. The exponentially decreasing density profile on the rear side of the target results in an electrostatic field or target-normal-sheath-acceleration field, $E_{\rm TNSA}$ \cite{Kumar2019a}. This TNSA field occurs in the upstream region and results in the upstream ions moving at velocity $v_0$ in the longitudinal direction \cite{Grismayer2006}.
%

To accelerate the ions via the CSA mechanism, the potential energy at the collisionless shock must exceed the kinetic energy of the upstream expanding ions. In other words, the electrostatic potential $\phi$ at the shock front satisfies the following condition \cite{Tidman}: $Z_i$$e\phi \geq \frac{1}{2}$$A_i$$m_p(V_{sh}^i-v_0^i)^2$, where $e$ is the electric charge, $A_i$ is the ion mass number, $m_p$ is the proton mass, $V^i_{sh}$ is the shock velocity, and the superscript $i$ represents the different ion species. The lower ion-velocity threshold ($v_L^i$) for ion reflection and CSA is $v_L^i = V_{sh}^i - \sqrt{2(Z_i /A_i)} e\phi / m_p$ \cite{Kumar2019a}.
Therefore, CSA occurs for
\begin{equation}
\label{eq:2}
v_L^i \leq v_0^i \leq V_{sh}^i.
\end{equation}
Equation (\ref{eq:2}) represents the lower $v_L^i$ and upper $V_{sh}^i$ bounds in $v_{0}^i$ for ion reflection. All ions with velocities $v_{0}^i$  between $v_L^i$ and $V_{sh}^i$ are reflected at the collisionless shock and leave with velocity 2$V_{sh}^i-v_0^i$ and the maximum velocity is $V_{sh}^i +  \sqrt{2(Z_i /A_i) e \phi / m_p}$ = 2$V_{sh}^i-v_L^i$.
For protons $Z_i = A_i = 1$, and the lower threshold is $v_L^{\rm P}  = V_{sh}^{\rm P}  - \sqrt{2e\phi /m_p}$.

\section{Results}
\subsection{Double-shock formation}
Figure \ref{fig:11} shows the phase-space of protons and C$^{6+}$ ions at $a_0$ = 3.35. A significant population of protons satisfy Eq. (\ref{eq:2}) and as result are accelerated at the collisionless shock [Fig. \ref{fig:11}(a)].
In comparison, relatively few C$^{6+}$ ions are reflected by the same collisionless shock, as this requires $V_{sh}^{\rm C}$ = $V_{sh}^{\rm P}$ [Fig. \ref{fig:11}(b)].
The lower threshold velocity for carbon ions, $v_L^{\rm C}$ is slightly larger than $v_L^{\rm P}$ as the charge-to-mass ratio, $Z_i/A_i$, is a factor of two smaller for C$^{6+}$.
Furthermore, because of the smaller $Z_i/A_i$, the expansion velocity $v_0^{\rm C}$ driven by $E_{\rm TNSA}$ in the upstream region is lower than equivalent process for protons. This causes $v_0^{\rm C}$ to drop below $v_L^{\rm C}$. This is illustrated in Fig. \ref{fig:11}(b) which highlights how few C$^{6+}$ ions are accelerated.
Indeed, some of the energetic C$^{6+}$ ions in Fig. \ref{fig:11}(b) likely originate early in time from the laser interaction at the front surface of the plasma. We conclude that a negligible number of C$^{6+}$ ions are accelerated via the CSA mechanism for $a_0 = 3.35$.

\begin{figure}[b]
  \includegraphics[width=0.49\textwidth]{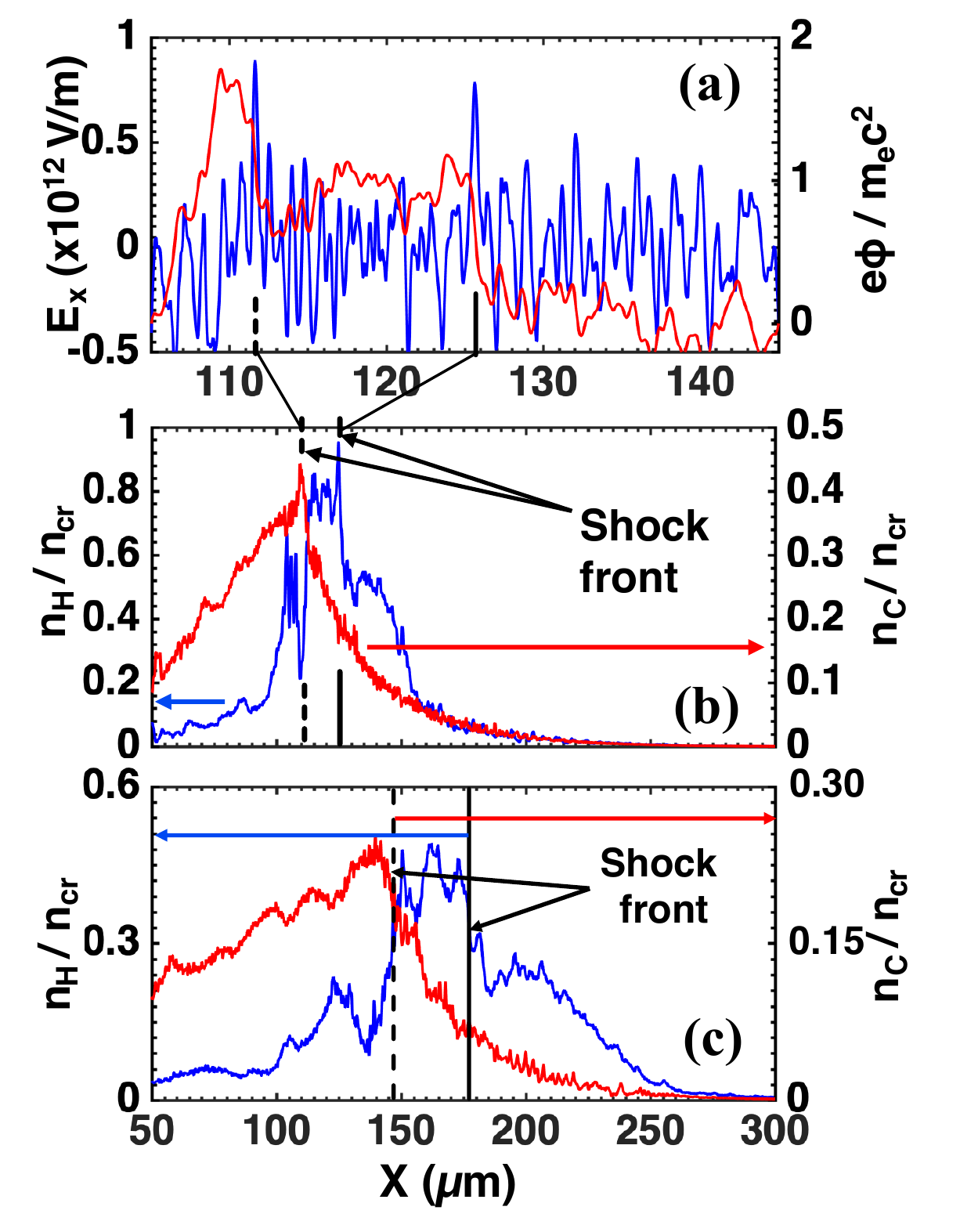}
  \caption{(a) The electrostatic field $E_x$ (left axis, blue line) and potential $\phi$ (right axis, red line) at $t = 3.0 ~\rm ps$. The normalized proton $n_H/n_{cr}$ (left axis, blue line) and carbon $n_{\rm C}/n_{cr}$ (right axis, red line) densities at (b) $t = 3.0 ~\rm ps$ and (c) $t = 4.0 ~\rm ps$ in $\rm C_2H_3Cl$ plasma for $a_0$ = 10. (a) is shown across a narrow longitudinal range compared to (b) and (c).}
   \label{fig:17}
\end{figure}

Simulations at low-intensity ($a_0 = 3.35$) generate a single-shock.
At higher intensity,  a key finding is the appearance of two distinct collisionless shocks. Figure \ref{fig:17} shows results at $a_0 = 10$, and Fig. \ref{fig:17}(a) illustrates the longitudinal electrostatic field $E_x$, averaged over the \textit{y}-axis, and potential $\phi$ at $t = 3.0 ~\mathrm{ps}$. Large amplitude changes in $E_x$ and $\phi$ are present at two different longitudinal positions. Large changes in the normalized proton and C$^{6+}$ ion densities are indicated, respectively, by the dotted ($x \approx 112 ~\rm \mu m$) and solid ($x \approx 126 ~\rm \mu m$) vertical lines in Fig. \ref{fig:17}(b). Figure \ref{fig:17}(c) shows how the normalized ion populations have evolved 1 ps later at $t = 4.0 ~\rm ps$. 
It is clear that the position of the jump in proton and C$^{6+}$ ion densities are different.
%
%

\begin{figure*}
  \includegraphics[width=0.9\textwidth]{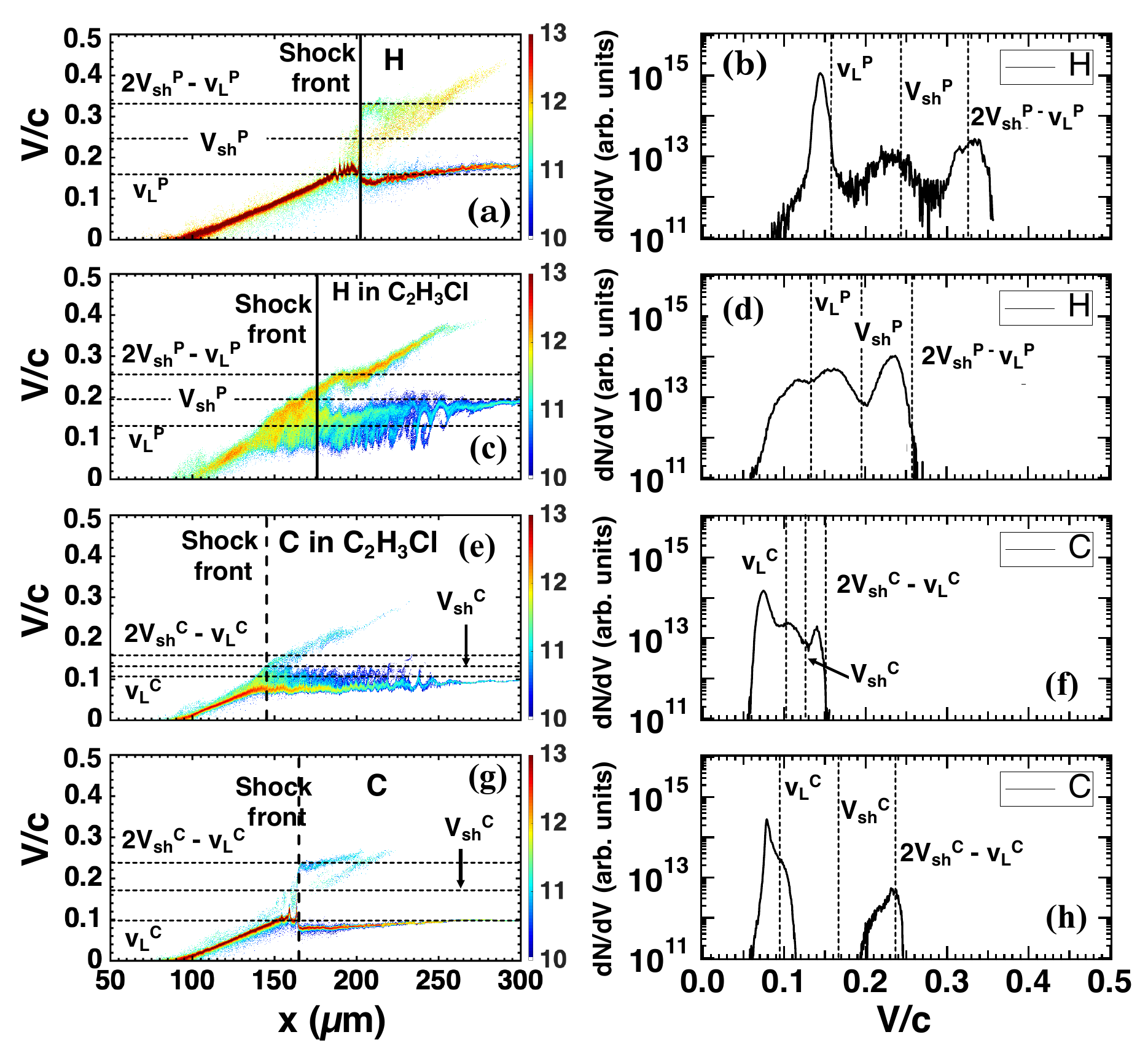}
  \caption{The first and second columns indicate the ion phase-space and the velocity spectra respectively for $a_0 = 10$ and at $t = 4.0 ~\rm ps$. The velocity spectra are taken in the upstream region immediately in front of the shock across $\Delta x = 3 ~\rm \mu m$. (a), (b) Results for protons from a  single-component H plasma. (c), (d) Results for protons from a $\rm C_2H_3Cl$ plasma. (e), (f) Results for C$^{6+}$ ions from a $\rm C_2H_3Cl$ plasma; and (g), (h) C$^{6+}$ ions in  single-component C plasma. The vertical lines on phase-space in panels (a), (c), (e), and (g) identify the position of the shock front.  In panels (b), (d), (f) and (h), moving left to right, the dotted lines indicate the positions of the lower threshold velocity ($v_L^i$), shock velocity ($V_{sh}^i$), and the maximum velocity of the reflected ions ($2V_{sh}^i-v_L^i$).  The color scale shows the number of ions in a log scale.}
   \label{fig:19}
\end{figure*}

Multiple shock structures are seen in the phase-space and velocity spectra in the first and second columns, respectively, of Fig.~\ref{fig:19}. The three sets of data are for single-component H, single-component C, and multicomponent $\rm C_2H_3Cl$ plasmas at $t = 4.0 ~\rm ps$. The positions of the shock fronts highlighted in Figs.~\ref{fig:19}(c) and \ref{fig:19}(e) are at the same longitudinal locations as the jumps in $n_{\rm H}/n_{cr}$ and $n_{\rm C} / n_{cr}$ identified in Fig.~\ref{fig:17}(c). 
A large number of protons and some of the C$^{6+}$ ions have velocities greater than $v_L^i$ and so CSA increases the velocity of these ions to $2V_{sh}^i-v_L^i$.
In the $\rm C_2H_3Cl$ plasma, collisionless shocks associated with the protons and separately with the C$^{6+}$ ions accelerate  the protons and C$^{6+}$ ions to different velocities.
Figures \ref{fig:19}(c) and \ref{fig:19}(e) indicate that the multicomponent $\rm C_2H_3Cl$ plasma develops, in the expanding upstream, a broad velocity distribution within the proton and C$^{6+}$ ion populations.
This is driven by an electrostatic ion two-stream instability (EITI) that arises from the velocity differences between the proton population with $Z_i/A_i=1$, and the heavier C$^{6+}$ ions with  $Z_i/A_i = 0.5$. We refer to this as heavy-ion EITI or HI-EITI \cite{Kumar2019a}. 
We find that the HI-EITI decelerates some upstream protons, while it accelerates some C$^{6+}$ ions with velocities below $v_L^{\rm C}$ to velocities that exceed this lower threshold, and thereby  increases the population of C$^{6+}$ ions available for CSA \cite{Kumar2019a}.
Furthermore, the CSA reflected-ion population, which moves at high velocity, causes an additional EITI with the slower moving expanding plasma that forms the upstream. We refer to this as reflected-ion EITI or RI-EITI \cite{Kumar2019a}. Overall, RI-EITI accelerates the slower upstream expanding ions towards higher velocity and promotes some ions, both protons and  C$^{6+}$ ions, with velocities below $v_L^i$ to velocities above the lower threshold. This further increases the ion population available for CSA \cite{Kumar2019a}. In a multicomponent $\rm C_2H_3Cl$ plasma, both RI-EITI and HI-EITI play essential roles in enabling the acceleration of C$^{6+}$ ions.
%
%
%
 \begin{figure*}
  \includegraphics[width=0.9\textwidth]{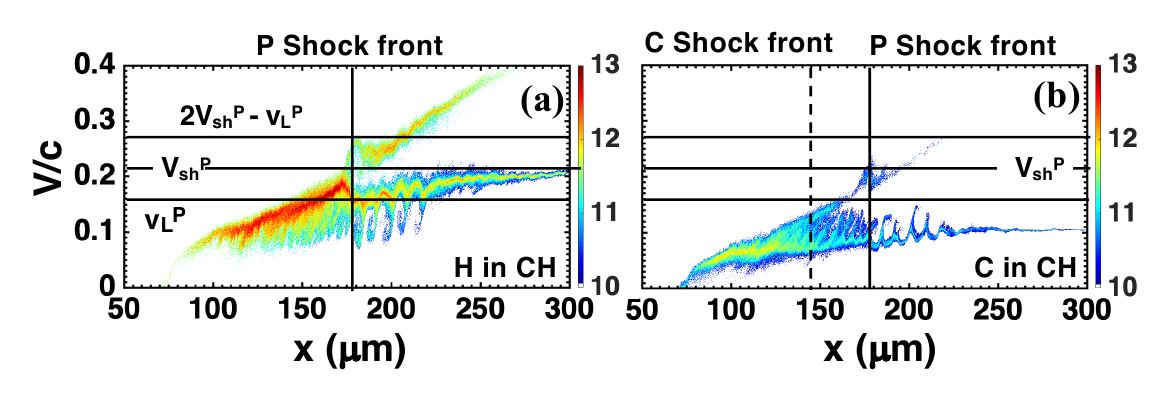}
  \caption{The phase-space for (a) protons and (b) C$^{6+}$ ions from a $\rm CH$ plasma for $a_0 = 10$ and at $t = 4.0 ~\rm ps$. The vertical lines identify the positions of the shock front associated with the protons (solid line) and the C$^{6+}$ ions (dashed line). The horizontal lines indicate the positions of the lower threshold velocity ($v_L^{\rm P}$) and shock velocity ($V_{sh}^{\rm P}$) of protons, and the maximum velocity of the reflected protons ($2V_{sh}^{\rm P}-v_L^{\rm P}$).} The color scale shows the number of ions in a log scale.
   \label{fig:CH}
\end{figure*}
\subsection{Double-step shock acceleration}
In the previous Section, the multiple-shock (proton shock and C$^{6+}$-ion shock) formation is described in a multicomponent $\rm C_2H_3Cl$ plasma at $a_0 = 10$, and protons and C$^{6+}$ ions are reflected and accelerated by each shock once. In this Section, we illustrate C$^{6+}$-ion acceleration is a double-step process with reflections at each shock in a multicomponent $\rm CH$ plasma at $a_0 = 10$.

Figures \ref{fig:CH} show the phase-space for protons [Fig.~\ref{fig:CH}(a)] and C$^{6+}$ ions [Fig.~\ref{fig:CH}(b)] in a $\rm CH$ plasma at $a_0 = 10$ and $t = 4.0~\rm ps$. We see that in this $\rm CH$ plasma, the high-mass C$^{6+}$ ions are reflected and accelerated twice; first at the C$^{6+}$ ion-shock ($x \approx 143 ~\rm \mu m$) and second at the proton-shock ($x \approx 177 ~\rm \mu m$) to the velocity $V/c = 0.23$. This is a clear observation of double-step multiple-shock acceleration of high-mass C$^{6+}$ ions in a multicomponent plasma.
This double-step shock acceleration of C$^{6+}$ ions is clearly seen in Fig.~\ref{fig:CH} but not in a $\rm C_2H_3Cl$ plasma [Fig.~\ref{fig:19}]. This is caused by a slightly faster proton-shock velocity of $V_{sh}^{\rm P}/c=0.22$ in a $\rm CH$ plasma compared with $V_{sh}^{\rm P}/c=0.20$ in a $\rm C_2H_3Cl$ plasma.  As a result, in a $\rm CH$ plasma $V_{sh}^{\rm P}$ is larger than the velocity of the pre-accelerated C$^{6+}$ ions, which are  reflected and accelerated by the C$^{6+}$-ion shock and likely originated from the laser interaction at the front surface of the plasma early in time. This results in the second acceleration of C$^{6+}$ ions by the proton-shock. In the case of a $\rm C_2H_3Cl$ plasma, $V_{sh}^{\rm P}$ is nearly equal to the velocity of the pre-accelerated C$^{6+}$ ions, and the second acceleration of C$^{6+}$ ions is not observed.

Furthermore, the respective deceleration and acceleration of expanding proton and C$^{6+}$ ion populations, as a result of HI-EITI, are more apparent in a $\rm CH$ plasma compared with a $\rm C_2H_3Cl$ plasma.
\subsection{The $a_0$ dependence of plasma parameters}
Simulations show the formation of two collisionless shocks at $a_0 \ge10$, and CSA of a significant number of C$^{6+}$ ions in multicomponent plasmas. This is qualitatively different from simulations at $a_0 = 3.35$ which show, see in Fig. \ref{fig:11}(b), a single shock. To understand the importance of increasing $a_0$, we extend our numerical investigation of CSA to $a_0 = 20$ and $33$ in a $\rm C_2H_3Cl$ plasma. These simulations confirm the existence of two collisionless shocks and indicate that the Mach number depends on $a_0$.

\begin{figure}
 \includegraphics[width=0.49\textwidth]{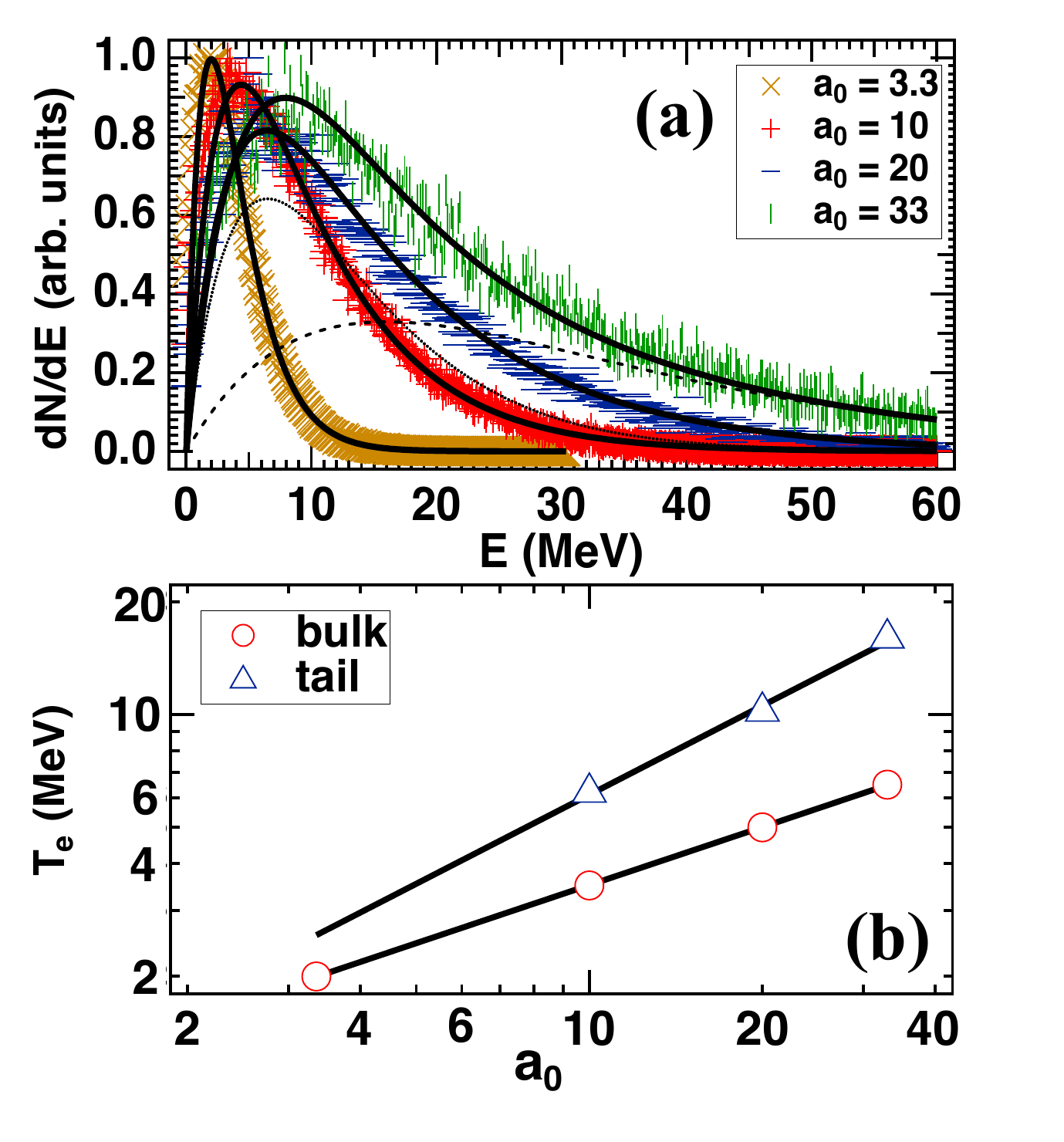}
\caption{(a) The electron energy distribution taken at $t = 4.0 ~\rm ps$ in the upstream region of the shock front for different laser intensities corresponding to $a_0 = 3.35$ ($\times$), $10$ (+), $20$ (-), and $33$ ($|$) for $\rm C_2H_3Cl$. A sum of two (bulk and tail) 2D relativistic Maxwellian is used to fit to the electron energy distribution shown by the solid lines for $a_0 = 10$, $20$, and $33$. The bulk (dotted line) and tail (dashed line) components for $a_0 = 33$ are shown. (b) The electron temperatures as a function of $a_0$.}
   \label{fig:20}
\end{figure}

In Fig. \ref{fig:20}(a) we compare, at $t = 4.0 ~\rm ps$, the upstream electron energy distributions for different $a_0$ and fit these with two-dimensional-relativistic (2D-relativistic) Maxwellian functions. The distributions at $a_0 = 10$, $20$, and $33$ are described by a two-temperature fit representing a bulk population and an energetic tail, while at  $a_0$ = 3.35 the distribution is described by a single temperature.
The bulk and tail Maxwellian components are shown for $a_0 = 33$.
Figure \ref{fig:20}(b) shows an $a_0$ power-law dependence for temperatures associated with the bulk and high-energy parts of the electron distributions.
The fitted electron temperatures do not depend on the target material, as the laser intensity and the electron densities are not material dependent but determined by $a_0$ \cite{Kumar2019a}.
%
%
%

The shock velocity $V_{sh}^i$ and the mean velocity $v_m^i$ of the expanding ions for all values of $a_0$ are higher in single-component H and C plasmas, compared with a multicomponent $\rm C_2H_3Cl$ plasma.
Furthermore, the difference between $V_{sh}^i$ and $v_{m}^i$, $v_{df}^i = V_{sh}^i - v_m^i$, increases with $a_0$ as a power-law except at the highest intensity, where $a_0$ = 33, which results from significant-levels ion reflection depleting or dissipating the collisionless shock \cite{Liseykina2015a}.
%
 \begin{figure}
  \includegraphics[width=0.49\textwidth]{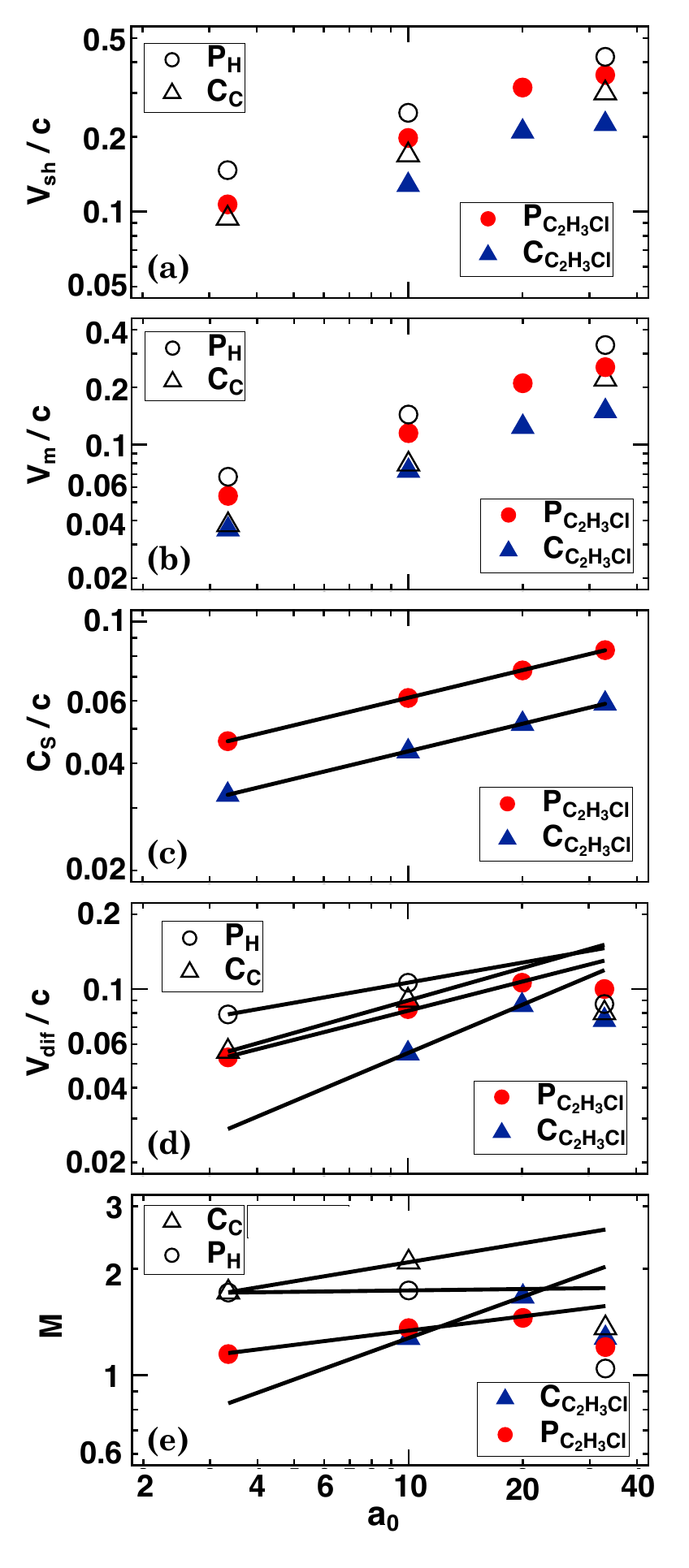}
  \caption{The $a_0$ dependence of (a) shock velocities $V_{sh}^i$, (b) mean velocities of the expanding ions $v_m^i$, (c) ion-acoustic velocities $c_s^i$, (d) difference between the shock velocity and mean velocity of the expanding ions $v_{df}^i= V_{sh}^i - v_m^i$, and (e) the corresponding Mach number $M^i = v_{df}^i/c_s^i$ at \textit{t} = 4.0 ps for protons in a single-component H plasma ($\bigcirc$), C$^{6+}$ ions in a single-component C plasma ($\triangle$), and protons (\textcolor{red}{\Huge $_\bullet$}) and C$^{6+}$ ions (\textcolor{blue}{\large $\blacktriangle$}) in a multicomponent $\rm C_2H_3Cl$ plasma.}
  \label{fig:21}
\end{figure}

Figures \ref{fig:21}(a) and \ref{fig:21}(b) represent the shock velocity ($V^i_{sh}$) and the mean velocity ($v^i_{m}$), respectively, of the expanding protons and C$^{6+}$ ions as a function of $a_0$ in a single-component H plasma, single-component C plasma, and $\rm C_2H_3Cl$ plasmas.
The proton and C$^{6+}$-ion $V_{sh}^i$ and $v_m^i$ are always larger for the single-component H plasma and single-component C plasma compared with the multicomponent $\rm C_2H_3Cl$ plasma, and follow the trend $V^{\rm P}_{\rm H}>V^{\rm P}_{\rm C_2H_3Cl}>V^{\rm C}_{\rm C}>V^{\rm C}_{\rm C_2H_3Cl}$ for all laser intensities. Here superscripts {P} and {C} denote protons and C$^{6+}$ ions, respectively, with different plasmas indicated by subscripts.
The ordering of velocities results from differences in the average charge-to-mass ratio $\langle Z \rangle / \langle A \rangle$, that is as $V^i_{sh}$ and $v^i_{m}$ are predominantly determined by the ion-acoustic velocity ($c^i_{s}$) and velocity of ions due to $E_{\rm TNSA}$, respectively.
Differences in the hole-boring velocity, which depends on $\sqrt{\langle Z \rangle / \langle A \rangle}$, explains why $V^{\rm P}_{\rm H}>V^{\rm P}_{\rm C_2H_3Cl}$ and $V^{\rm C}_{\rm C}>V^{\rm C}_{\rm C_2H_3Cl}$ \cite{Kumar2019a}.
As a result, the shock velocity in a single-component H plasma (with $\langle Z \rangle / \langle A \rangle$ = 1) is larger than that in $\rm C_2H_3Cl$  ($\langle Z \rangle / \langle A \rangle$ = 0.48), and shock velocity for C$^{6+}$ ion in a single-component C plasma ($\langle Z \rangle / \langle A \rangle$ = 0.50) is larger than that in $\rm C_2H_3Cl$.

Ion-acoustic waves are excited in proton and C$^{6+}$ ion populations and using the bulk electron temperatures $T_e$  to derive an ion-acoustic velocity, $c_s^i = \sqrt{(Z_i/A_i)T_e/m_p}$, we find that the associated Mach numbers, $M^i = v_{df}^i/c_s^i$, scale as a power law in $a_0$.
The ion-acoustic velocities for protons ($c_s^{\rm P}$) and C$^{6+}$ ions  ($c_s^{\rm C}$) are calculated using the bulk temperature of the plasma. The upstream bulk temperatures in a single-component H plasma, single-component C plasma, and multicomponent $\rm C_2H_3Cl$ plasma are the same, as a result the $c_s^i$  depends on the  $\sqrt{\langle Z \rangle / \langle A \rangle}$.
This is shown in Fig.~\ref{fig:21}(c). The $c_s^i$, indicated by the solid lines, scale with $a_0$ as a power-law. The difference between the shock velocity and mean velocity of the expanding ions, i.e., $v_{df}^i = V_{sh}^i - v_m^i$, is shown in Fig.~\ref{fig:21}(d) and increases as a power-law with $a_0$ except at $a_0$ = 33.

The ratio between the $v_{df}^i$ and $c_s^i$ yields the Mach number $M^i = v_{df}^i/c_s^i$, this is shown in Fig.~\ref{fig:21}(e). In comparison with $M^{\rm P}$, the Mach number for protons, $M^{\rm C}$, the Mach number for C$^{6+}$ ions, has a strong scaling with $a_0$. Notice in multicomponent $\rm C_2 H_3Cl$, $M^{\rm C} <1 $ for $a_0$ = 3.35, and no shock is associated with the C$^{6+}$ ions.
Furthermore,  $M^{\rm P}$ in a single-component H plasma decreases with $a_0$, this occurs as $v_m^{\rm P}$ scales faster with $a_0$ than $V_{sh}^{\rm P}$, causing a slow scaling of $v_{df}^{\rm P}$ compared with $c_s^{\rm P}$ as $a_0$ increases.

\begin{figure}
  \includegraphics[width=0.49\textwidth]{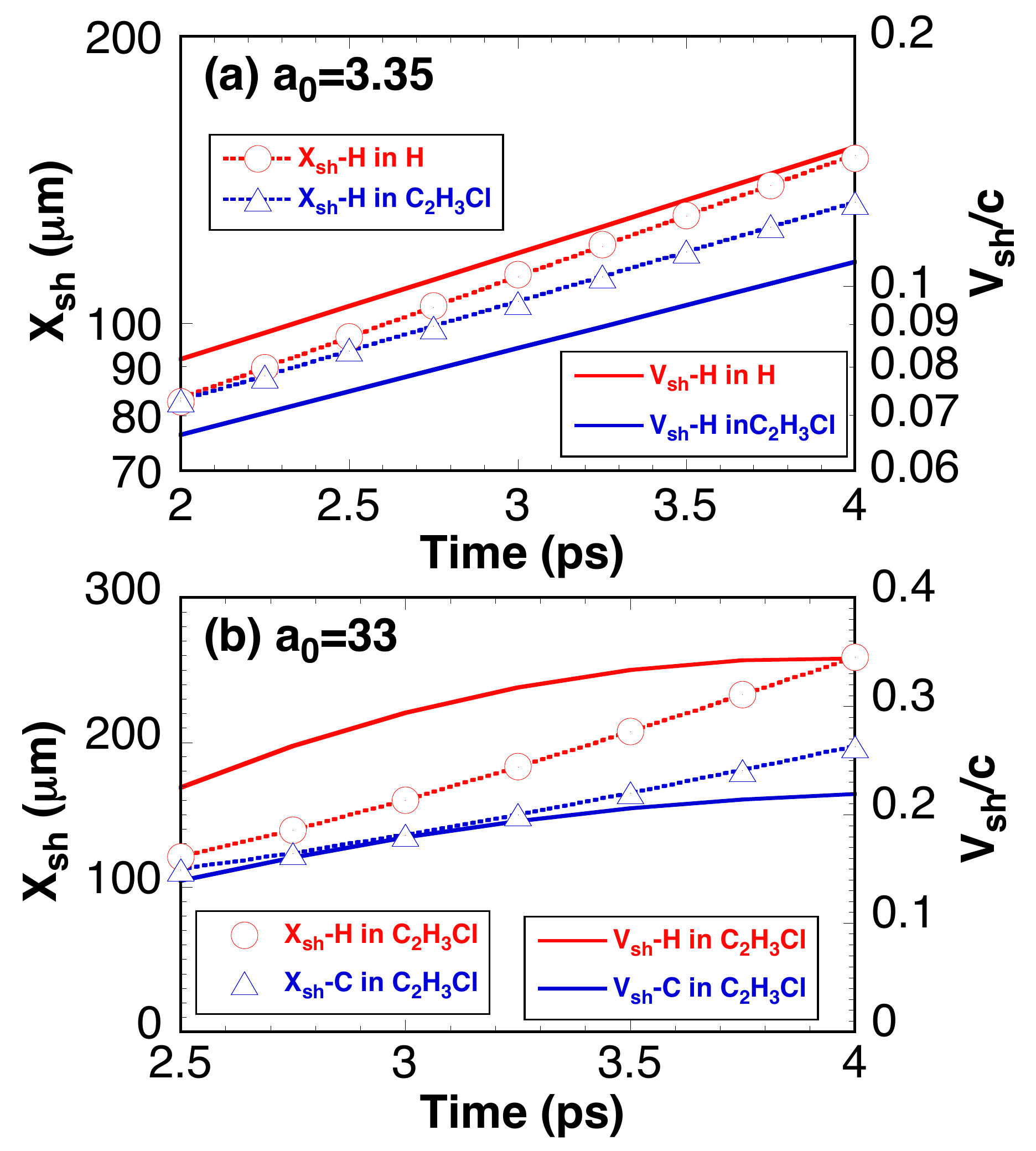}
  \caption{(a) The temporal evolution of shock positions $X_{sh}$ and shock velocities $V_{sh}$ for protons in a single-component H plasma (shown in red) and a multicomponent $\rm C_2H_3Cl$ plasma (shown in blue) for $a_0$ = 3.35. $X_{sh}$ data are shown as open circles (a single-component H plasma:\textcolor{red}{$\bigcirc$}) and open triangles (a multicomponent $\rm C_2H_3Cl$ plasma:\textcolor{blue}{$\triangle$}). The derivative of $X_{sh}$ with respect to time gives $V_{sh}$.  $X_{sh}$ (dotted lines) and $V_{sh}$ (solid lines) rise exponentially with time.
(b) The temporal evolution of $X_{sh}$ and $V_{sh}$ for protons (shown in red) and C$^{6+}$ ions (shown in blue) in a multicomponent $\rm C_2H_3Cl$ plasma for $a_0$ = 33. $X_{sh}$ data are shown as open circles (protons:\textcolor{red}{$\bigcirc$}) and open triangles (C$^{6+}$ ions:\textcolor{blue}{$\triangle$}). The time dependencies of $X_{sh}$ and $V_{sh}$ are best represented by a third-order (dotted lines) and second-order (solid lines) order polynomials, respectively.}
   \label{fig:Vsh}
\end{figure}

\subsection{Shock dissipation}
For $a_0$ = 33 shock dissipation, driven by ion reflection, becomes more pronounced. This reduces the shock velocity \cite{Liseykina2015a}. Evidence for this is seen in Fig.~\ref{fig:21}(d) of $v_{df}^i$ and in Fig.~\ref{fig:21}(e) of $M^i$ which illustrate a power-law trend for $a_0 = 3.35$, $10$, and $20$ up to the end of the simulations at $t=4.0$ ps. Simulations at $a_0= 33$ show significant shock dissipation from $t \approx 2.5$ ps.
For $a_0 = 3.35$, $10$, and $20$, shock velocities increase exponentially with time until $t = 4.0 ~\rm ps$, in contrast, for $a_0 = 33$, the shock velocity increases to $t < 2.5 ~\rm ps$ then dissipates, which results in low Mach numbers at $a_0 = 33$ for single- and multicomponent plasmas \cite{Liseykina2015a}.

The temporal variation of shock positions ($X_{sh}$) in a single-component H plasma and a multicomponent $\rm C_2H_3Cl$ plasma for $a_0$ = 3.35 are shown in Fig.~\ref{fig:Vsh}(a). The derivative $dX_{sh}/dt$ gives the shock velocity $V_{sh}$.
Since there is an exponential drop in the density at the rear-side of the target, $V_{sh}$ increases exponentially as a function of time for all target materials. For $a_0$ = 3.35 [see Fig.~\ref{fig:Vsh}(a)], 10, and 20, $X_{sh}$ and $V_{sh}$ rise exponentially with time.
Comparing this to Fig.~\ref{fig:Vsh}(b), we see that at $a_0$ = 33 the temporal evolution of $X_{sh}$ and $V_{sh}$ for protons and C$^{6+}$ ions in a multicomponent $\rm C_2H_3Cl$ plasma is slower. Indeed, the time dependencies of $X_{sh}$ and $V_{sh}$ are best represented by third- and second-order polynomials, respectively. This slower temporal evolution results from enhanced ion reflection at the shock which increases shock dissipation \cite{Liseykina2015a}.

 \begin{figure}
  \includegraphics[width=0.49\textwidth]{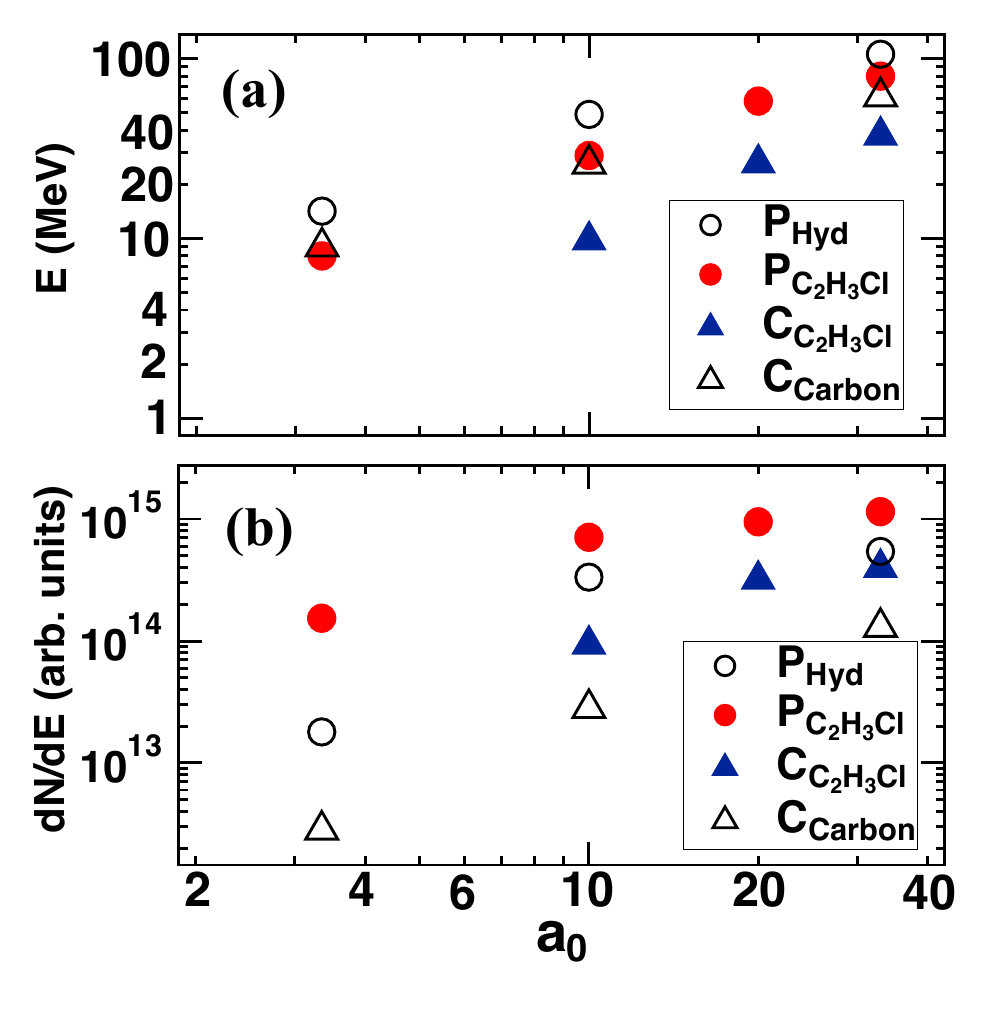}
  \caption{The $a_0$ dependence of (a) energy $E$ per nucleon and (b) the number $dN/dE$ at $E$ of reflected protons and C$^{6+}$ ions at the peak of the energy distribution at \textit{t} = 4.0 ps for protons in  single-component H ($\bigcirc$), C$^{6+}$ ions in  single-component C ($\triangle$), and protons (\textcolor{red}{\Huge $_\bullet$}) and C$^{6+}$ ions (\textcolor{blue}{\large $\blacktriangle$}) in $\rm C_2H_3Cl$ plasmas.}
   \label{fig:22}
\end{figure}
%

%

\subsection{The $a_0$ dependence of ion acceleration}
Figures \ref{fig:22}(a) and \ref{fig:22}(b) show how the energy $E$ and the number of reflected ions at the peak of the energy distribution $dN/dE$ depend on $a_0$.
In the $\rm C_2H_3Cl$ plasma there are no C$^{6+}$-associated shocks at $a_0 = 3.35$ as $M^{\rm C}<1$.
The energies of the reflected ions are always larger in single-component H or C plasma when compared with multicomponent $\rm C_2H_3Cl$ plasma [Fig. \ref{fig:22}(a)]. 
This is a feature of smaller $V_{sh}^i$ and amplitude of $\phi$ in the multicomponent $\rm C_2H_3Cl$ compared with the single-component H or C plasma as shown in Fig. \ref{fig:19}.

\begin{figure}[h]
  \includegraphics[width=0.49\textwidth]{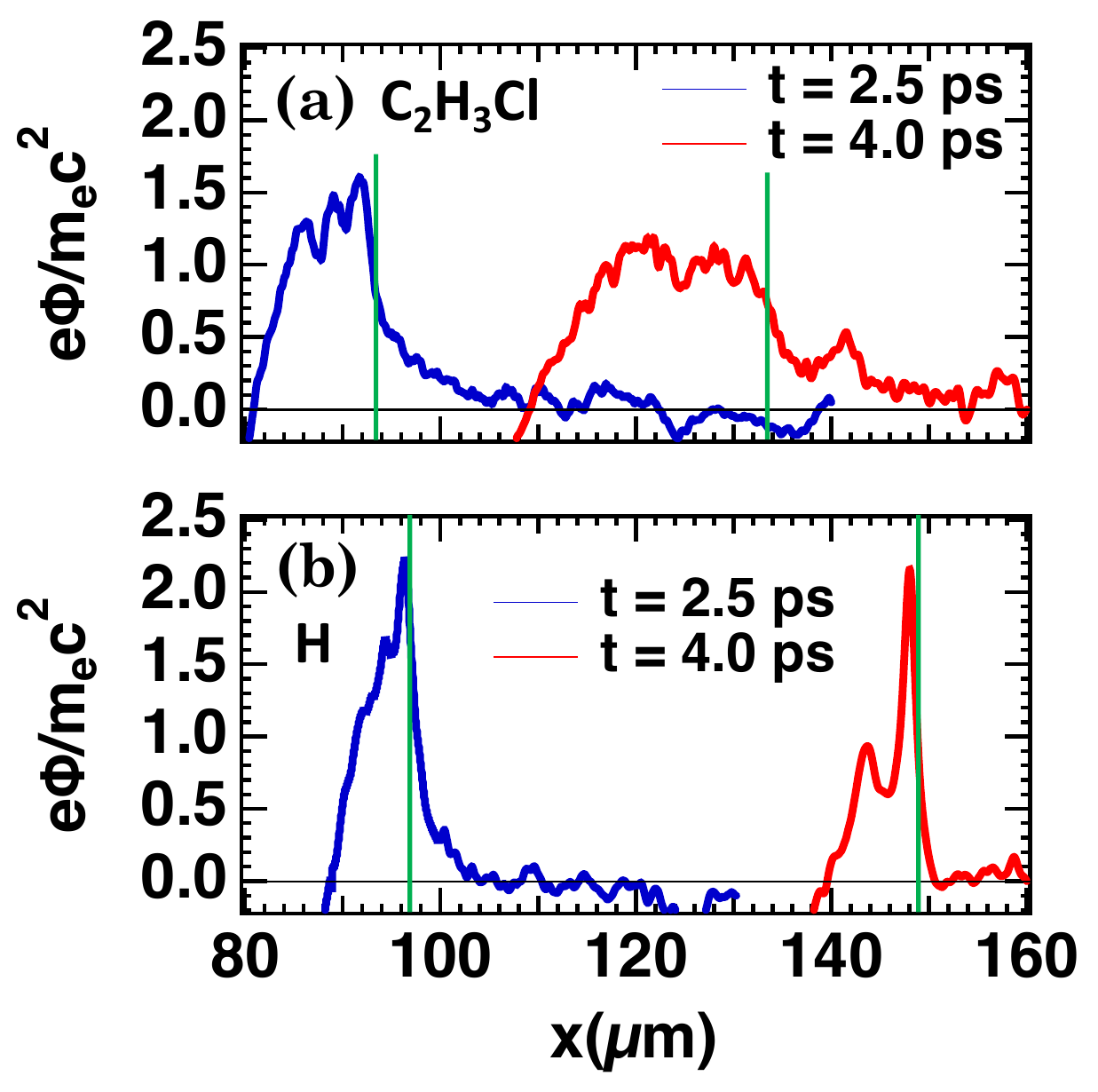}
  \caption{The spatial profile of electrostatic potentials in (a) a multicomponent $\rm C_2H_3Cl$ plasma and (b) a single-component hydrogen plasma at $t = 2.5$ (blue curves) and 4.0 ps (red curves) for $a_0=3.35$. The vertical lines indicate the position of the shock fronts.}
   \label{fig:potential}
\end{figure}

Figures \ref{fig:potential}(a) and \ref{fig:potential}(b) show the spatial profile of electrostatic potentials $\phi$ in a multicomponent $\rm C_2H_3Cl$ plasma and a single-component H plasma, respectively, at $t =$ 2.5 (blue curve) and 4.0 ps (red curve) for $a_0=3.35$. 
The vertical lines indicate the position of the shock fronts. These highlight that $\phi$ is smaller in a multicomponent $\rm C_2H_3Cl$ plasma compared to a single-component H plasma.
The smaller $\phi$ is a result of a lower $\langle Z \rangle / \langle A \rangle$ plasma.
In a single-component H plasma, the gradient $d\phi/dx$ is large. This potential jump is associated with the shock and is necessary for ion acceleration. It is produced by a charge separation between electrons and ions.  
This feature is not observed in the multicomponent $\rm C_2H_3Cl$ plasma because the charge separation between electrons and ions is smeared out over a larger volume by the heavier C and Cl ions, as a result the amplitude and gradient associated with $\phi$ are smaller.
 
Our PIC results indicate that $V_{sh}$ is smaller at lower $\langle Z \rangle / \langle A \rangle$, this is explained by recognising that the hole-boring velocity \cite{Robinson2009} 
\begin{equation}
\label{eq:3}
V_{HB} = c \sqrt{a_0^2 \frac{\langle Z \rangle}{\langle A \rangle} \frac{m_e}{m_p} \frac{n_{\rm cr}}{n_e}},
\end{equation}
determines the velocity of the piston driving the collisionless shock. 
Given that $a_0$ and $n_e$ are same for all target materials, $V_{HB}$ has relative dependence on only $\sqrt{\langle Z \rangle / \langle A \rangle}$, maximizing $V_{sh}$ when $\langle Z \rangle / \langle A \rangle$ is largest, i.e., for a single-component H plasma.
%
%

For the multicomponent $\rm C_2H_3Cl$ plasma at $a_0 > 3.35$ the flux of the reflected protons and C$^{6+}$ ions is higher [see Fig.~\ref{fig:22}(b)].
It is important to note that more protons are accelerated in multicomponent plasma as $v_L^{\rm P}$ is lower in comparison with the single-component H plasma.
For C$^{6+}$ ion acceleration,  HI-EITI, which is present only in a multicomponent plasma, broadens the velocity distribution of the expanding C$^{6+}$ ions towards higher velocity. This results in more C$^{6+}$ ions being available for CSA in comparison to single-component C plasmas.
These results confirm our earlier observation \cite{Kumar2019a} that only proton collisionless shocks were observed in multicomponent plasmas at $a_0$ = 3.35.
In this work Mach numbers $M$ = 1.6 - 1.7 were calculated with a critical Mach number needed for the proton reflection and CSA. These values were derived using ion-acoustic velocities based on a $\langle Z \rangle / \langle A \rangle$, where $\langle Z \rangle$ and $\langle A \rangle$ are the respective averages of $\langle Z_i \rangle$ and $\langle A_i \rangle$ across all ion species in a plasma. Here, we use ion-specific ion-acoustic velocities to describe the two collisionless shocks. For protons $c_s^{\rm P}$ determines the Mach number $M^{\rm P}$ of a proton collisionless shock, and ions satisfying the reflection condition given by Eq. (\ref{eq:2}) are accelerated even when the Mach number is less than $M_{\rm cr}$ defined in Ref.~\cite{Kumar2019a}.
%
%

%
\section{Summary}
Two-dimensional PIC simulations are used to investigate the evolution of electrostatic collisionless shocks and CSA of protons and heavy ions in multicomponent plasmas.
The interaction of a high-intensity p-polarized laser with $\rm C_2H_3Cl$ and $\rm CH$ plasmas leads to the formation of the two shock fronts in proton and C$^{6+}$-ion populations.
Both shocks have different amplitudes of the shock potential and propagate with different velocities.
The electron temperature, shock velocities, and Mach numbers for shocks associated with proton ($M^{\rm P}$) and C$^{6+}$ ions ($M^{\rm C}$) scale as a power-laws with the normalized laser intensity $a_0$. In the multicomponent $\rm C_2H_3Cl$ plasma, $M^{\rm C}$ scales faster with $a_0$ compared to $M^{\rm P}$. At $a_0 = 3.35$, as $M^{\rm C} < 1$, a C$^{6+}$ ion-shock does not form. On increasing $a_0$, shock formation with CSA of protons and C$^{6+}$ ions occurs at different location and velocities. 
Double-step shock acceleration is investigated in a $\rm CH$ plasma, in which the pre-accelerated C$^{6+}$ ions are further accelerated at the proton-shock.
A broadening upwards of the C$^{6+}$ ion velocity distribution, as a result of a HI-EITI, is important and increases the number of C$^{6+}$ ions accelerated.
For $a_0$ = 33 shock dissipation, driven by ion reflection, becomes more pronounced. This results in the reduction of the shock velocity.
Moreover, modern ultra-intense, picosecond duration lasers enable the laboratory study of the formation and modification of collisionless shocks as ions are accelerated in multicomponent plasmas. These topics are important to space physics, astrophysics, and plasma physics.

\section{Acknowledgement}
This research was partially supported by Japan Society for the Promotion of Science (JSPS) KAKENHI Grant No. JP15H02154, JP17H06202, JP19H00668, JSPS Core-to-Core Program B. Asia-Africa Science Platforms Grant No. JPJSCCB20190003, EPSRC grant EP/L01663X/1 and EP/P026796/1.
%


%
\end{document}